\documentclass[12pt,prd,showpacs,tightenlines,nofootinbib]{revtex4}
\usepackage{bm}
\usepackage{graphics}
\usepackage{rotating}
\usepackage{epsfig}
\begin{document}
\begin{flushright}{HU-EP-07/17}\end{flushright}
\title{Masses of excited heavy baryons in the relativistic quark-diquark
picture}
\author{D. Ebert$^1$, R. N. Faustov$^{1,2}$  and V. O. Galkin$^{1,2}$}
\affiliation{$^1$ Institut f\"ur Physik, Humboldt--Universit\"at zu Berlin,
Newtonstr. 15, D-12489  Berlin, Germany\\
$^2$ Dorodnicyn Computing Centre, Russian Academy of Sciences,
  Vavilov Str. 40, 119991 Moscow, Russia}

\begin{abstract}
The mass spectra of the excited heavy baryons consisting of two light
($u,d,s$) and one heavy ($c,b$) quarks are calculated in the 
heavy-quark--light-diquark approximation within the constituent quark
model. The light quarks, forming the diquark, and the light diquark in the
baryon are treated completely relativistically. The expansion in
$v/c$ up to the second order  is
used only for the heavy ($b$ and $c$) quarks. 
The internal structure of the diquark is taken into account by
inserting the diquark-gluon interaction form factor. An overall
good agreement of the obtained predictions with available
experimental data is found.       
 
\end{abstract}


\maketitle

In this letter we extend our previous study of the ground state masses
of heavy baryons \cite{hbar} to the description of their excited states. All
calculations are performed in the framework of the relativistic quark
model based on the quasipotential approach in quantum field theory. As
in our previous analysis we use the heavy-quark--light-diquark
approximation to reduce a very complicated relativistic three-body
problem to the subsequent solution of two more simple two-body
problems. The first step consists in the calculation of the masses, wave
functions and form factors of the diquarks, composed from two light
quarks. Next, at the second step, a heavy baryon is treated as a
relativistic bound system of a light diquark and heavy quark. It is
important to emphasize that we do not consider a diquark as a point
particle but explicitly take into account its structure by calculating
the diquark-gluon interaction form factor through the diquark wave
functions.  Such
scheme proved to be very effective and successful in our calculation
of the ground state masses of heavy baryons. The obtained results were
found to be in good agreement with experimental data \cite{pdg}. Moreover, the
predicted masses of the $\Omega_c^*$ and $\Sigma_b$, $\Sigma_b^*$ proved
to be very close to the recently measured ones
\cite{babaromega,cdfSigma}. This 
gives us additional confidence in the reliability of the used
quark-diquark approximation within our model. 

It is important to point out that during last few years a significant
experimental progress has been achieved in studying heavy baryons with
one heavy quark. At present all masses of ground states of charmed baryons as
well as many of their excitations are known experimentally with
rather good precision \cite{pdg}. Putting into operation the Large Hadron
Collider (LHC) will provide us with data on masses of excited bottom
baryons as well. Therefore the calculation of the mass spectra of excited
heavy baryons turns to be a really actual problem. Here we consider only the
orbital and radial excitations of the light diquark with respect to the
heavy quark. There are strong theoretical indications \cite{isg} that for
such excitations the quark-diquark approximation should work
even better than for the ground state heavy baryons.

In the quasipotential approach and quark-diquark picture of
heavy baryons the interaction of two light quarks in a diquark and the heavy
quark interaction with a light diquark in a baryon are described by the
diquark wave function ($\Psi_{d}$) of the bound quark-quark state
and by the baryon wave function ($\Psi_{B}$) of the bound quark-diquark
state respectively,  which satisfy the
quasipotential equation \cite{3} of the Schr\"odinger type \cite{4}
\begin{equation}
\label{quas}
{\left(\frac{b^2(M)}{2\mu_{R}}-\frac{{\bf
p}^2}{2\mu_{R}}\right)\Psi_{d,B}({\bf p})} =\int\frac{d^3 q}{(2\pi)^3}
 V({\bf p,q};M)\Psi_{d,B}({\bf q}),
\end{equation}
where the relativistic reduced mass is
\begin{equation}
\mu_{R}=\frac{M^4-(m^2_1-m^2_2)^2}{4M^3},
\end{equation}
and $E_1$, $E_2$ are given by
\begin{equation}
\label{ee}
E_1=\frac{M^2-m_2^2+m_1^2}{2M}, \quad E_2=\frac{M^2-m_1^2+m_2^2}{2M}.
\end{equation}
Here $M=E_1+E_2$ is the bound state mass (diquark or baryon),
$m_{1,2}$ are the masses of light quarks ($q_1$ and $q_2$) which form
the diquark or of the light diquark ($d$) and heavy quark ($Q$) which form
the heavy baryon ($B$), and ${\bf p}$  is their relative momentum.  
In the center of mass system the relative momentum squared on mass shell 
reads
\begin{equation}
{b^2(M) }
=\frac{[M^2-(m_1+m_2)^2][M^2-(m_1-m_2)^2]}{4M^2}.
\end{equation}

The kernel 
$V({\bf p,q};M)$ in Eq.~(\ref{quas}) is the quasipotential operator of
the quark-quark or quark-diquark interaction. It is constructed with
the help of the
off-mass-shell scattering amplitude, projected onto the positive
energy states. In the following analysis we closely follow the
similar construction of the quark-antiquark interaction in mesons
which were extensively studied in our relativistic quark model
\cite{egf}. For
the quark-quark interaction in a diquark we use the relation
$V_{qq}=V_{q\bar q}/2$ arising under the assumption about the octet
structure of the interaction  from the difference of the $qq$ and
$q\bar q$  colour states. An important role in this construction is
played by the Lorentz-structure of the nonperturbative confining  interaction. 
In our analysis of mesons, while  
constructing the quasipotential of the quark-antiquark interaction, 
we adopted that the effective
interaction is the sum of the usual one-gluon exchange term with the mixture
of long-range vector and scalar linear confining potentials, where
the vector confining potential contains the Pauli terms.  
We use the same conventions for the construction of the quark-quark
and quark-diquark interactions in the baryon. The
quasipotential  is then defined by the following expressions \cite{efgm,egf} 

(a) for the quark-quark ($qq$) interaction
 \begin{equation}
\label{qpot}
V({\bf p,q};M)=\bar{u}_{1}(p)\bar{u}_{2}(-p){\cal V}({\bf p}, {\bf
q};M)u_{1}(q)u_{2}(-q),
\end{equation}
with
\[
{\cal V}({\bf p,q};M)=\frac12\left[\frac43\alpha_sD_{ \mu\nu}({\bf
k})\gamma_1^{\mu}\gamma_2^{\nu}+ V^V_{\rm conf}({\bf k})
\Gamma_1^{\mu}({\bf k})\Gamma_{2;\mu}(-{\bf k})+
 V^S_{\rm conf}({\bf k})\right],
\]

(b) for quark-diquark ($Qd$) interaction
\begin{eqnarray}
\label{dpot}
V({\bf p,q};M)&=&\frac{\langle d(P)|J_{\mu}|d(Q)\rangle}
{2\sqrt{E_d(p)E_d(q)}} \bar{u}_{Q}(p)  
\frac43\alpha_sD_{ \mu\nu}({\bf 
k})\gamma^{\nu}u_{Q}(q)\cr
&&+\psi^*_d(P)\bar u_Q(p)J_{d;\mu}\Gamma_Q^\mu({\bf k})
V_{\rm conf}^V({\bf k})u_{Q}(q)\psi_d(Q)\cr 
&&+\psi^*_d(P)
\bar{u}_{Q}(p)V^S_{\rm conf}({\bf k})u_{Q}(q)\psi_d(Q), 
\end{eqnarray}
where $\alpha_s$ is the QCD coupling constant, $\langle
d(P)|J_{\mu}|d(Q)\rangle$ is the vertex of the 
diquark-gluon interaction which takes into account the diquark
internal structure. $D_{\mu\nu}$ is the  
gluon propagator in the Coulomb gauge, ${\bf k=p-q}$; $\gamma_{\mu}$ and $u(p)$ are 
the Dirac matrices and spinors.

The diquark state in the confining part of the quark-diquark
quasipotential (\ref{dpot}) is described by the wave functions
\begin{equation}
  \label{eq:ps}
  \psi_d(p)=\left\{\begin{array}{ll}1 &\qquad \text{ for scalar diquark}\\
\varepsilon_d(p) &\qquad \text{ for axial vector diquark}
\end{array}\right. ,
\end{equation}
where $\varepsilon_d$ is the polarization vector of the axial vector
diquark. The effective long-range vector vertex of the
diquark can be presented in the form  
\begin{equation}
  \label{eq:jc}
  J_{d;\mu}=\left\{\begin{array}{ll}
  \frac{\displaystyle (P+Q)_\mu}{\displaystyle
  2\sqrt{E_d(p)E_d(q)}}&\qquad \text{ for scalar diquark}\cr
\frac{\displaystyle (P+Q)_\mu}{\displaystyle2\sqrt{E_d(p)E_d(q)}}
  -\frac{\displaystyle i\mu_d}{\displaystyle 2M_d}\Sigma_\mu^\nu 
\tilde k_\nu
  &\qquad \text{ for axial 
  vector diquark}\end{array}\right. ,
\end{equation}
where $\tilde k=(0,{\bf k})$. Here the $\Sigma_\mu^\nu$ is the antisymmetric tensor
\begin{equation}
  \label{eq:Sig}
  \left(\Sigma_{\rho\sigma}\right)_\mu^\nu=-i(g_{\mu\rho}\delta^\nu_\sigma
  -g_{\mu\sigma}\delta^\nu_\rho),
\end{equation}
and the axial vector diquark spin ${\bf S}_d$ is given by
$(S_{d;k})_{il}=-i\varepsilon_{kil}$. We choose the total
chromomagnetic moment of the axial vector 
diquark $\mu_d=0$ \cite{tetr}.

The effective long-range vector vertex of the quark is
defined by \cite{egf,sch}
\begin{equation}
\Gamma_{\mu}({\bf k})=\gamma_{\mu}+
\frac{i\kappa}{2m}\sigma_{\mu\nu}\tilde k^{\nu}, \qquad \tilde
k=(0,{\bf k}),
\end{equation}
where $\kappa$ is the Pauli interaction constant characterizing the
anomalous chromomagnetic moment of quarks. In the configuration space
the vector and scalar confining potentials in the nonrelativistic
limit reduce to
\begin{eqnarray}
V^V_{\rm conf}(r)&=&(1-\varepsilon)V_{\rm conf}(r),\nonumber\\
V^S_{\rm conf}(r)& =&\varepsilon V_{\rm conf}(r),
\end{eqnarray}
with 
\begin{equation}
V_{\rm conf}(r)=V^S_{\rm conf}(r)+
V^V_{\rm conf}(r)=Ar+B,
\end{equation}
where $\varepsilon$ is the mixing coefficient.

The constituent quark masses $m_b=4.88$ GeV, $m_c=1.55$ GeV,
$m_u=m_d=0.33$ GeV, $m_s=0.5$ GeV and 
the parameters of the linear potential $A=0.18$ GeV$^2$ and $B=-0.3$ GeV
have the usual values of quark models.  The value of the mixing
coefficient of vector and scalar confining potentials $\varepsilon=-1$
has been determined from the consideration of charmonium radiative
decays \cite{efg} and the heavy quark expansion \cite{fg}. 
Finally, the universal Pauli interaction constant $\kappa=-1$ has been
fixed from the analysis of the fine splitting of heavy quarkonia ${
}^3P_J$- states \cite{efg}. In the literature the 't~Hooft-like
interaction between quarks induced by instantons \cite{dk} is widely
discussed. 
This interaction can be effectively described by introducing the quark
anomalous chromomagnetic moment having an approximate value
$\kappa\approx-0.75$ (Diakonov \cite{dk}). This value is of the same
sign and order of magnitude as the Pauli constant $\kappa=-1$ in our
model. Thus the Pauli term incorporates at least partly  the
instanton contribution to the $q\bar q$ interaction.  Note that the 
long-range chromomagnetic contribution to the potential in our model
is proportional to $(1+\kappa)$ and thus vanishes for the 
chosen value of $\kappa=-1$.

At the first step, we calculate the masses and form factors of the light
diquark. As it is well known, the light quarks are highly
relativistic, which makes the $v/c$ expansion inapplicable and thus,
a completely relativistic treatment is required. To achieve this goal in
describing light 
diquarks, we closely follow our recent consideration of the spectra of light
mesons \cite{lmes} and adopt the same procedure to make the relativistic
quark potential local by replacing
$\epsilon_{1,2}(p)=\sqrt{m_{1,2}^2+{\bf p}^2}\to E_{1,2}$  
(see (\ref{ee}) and discussion in Ref.~\cite{lmes}). 

The quasipotential equation (\ref{quas}) is solved numerically for the
complete relativistic potential  which depends on the
diquark mass in a complicated highly nonlinear way \cite{hbar}.  The obtained
ground state masses of scalar  and axial
vector  light diquarks are presented in
Table~\ref{tab:mass}. These  masses are in good agreement with values
found within the Nambu--Jona-Lasinio model \cite{efkr}, by solving  the
Bethe-Salpeter equation with different types of kernel
\cite{burden,maris} and in quenched lattice calculations \cite{hess}.
It follows from  Table~\ref{tab:mass} that the mass difference between
the scalar and vector diquark decreases from $\sim 200$ to $\sim 120$ MeV,
when one of the $u,d$ quarks is replaced by the $s$ quark in accord
with the statement of Ref.~\cite{jaffe}.    
\begin{table}
  \caption{Masses of light ground state diquarks (in MeV). S and A
    denotes scalar and axial vector diquarks antisymmetric $[q,q']$ and
    symmetric $\{q,q'\}$ in flavour, respectively. }
  \label{tab:mass}
\begin{ruledtabular}
\begin{tabular}{ccccccc}
Quark& Diquark&  
\multicolumn{5}{c}{\hspace{-2.9cm}\underline{\hspace{5.1cm}Mass\hspace{5.1cm}}}
\hspace{-2.9cm} \\
content &type & \cite{hbar}& \cite{efkr}&\cite{burden}&\cite{maris} &
\cite{hess}\\
& &our &NJL &BSE & BSE &Lattice\\
\hline
$[u,d]$& S & 710 & 705 &737 &820& 694(22)\\
$\{u,d\}$& A & 909 & 875 &949 &1020&806(50)\\
$[u,s]$ & S& 948 & 895 &882&1100&\\
$\{u,s\}$& A & 1069 & 1050&1050&1300& \\
$\{s,s\}$& A & 1203 & 1215&1130&1440& \\
\end{tabular}
\end{ruledtabular}
\end{table}
 
In order to determine the diquark interaction with the gluon field, which
takes into account the diquark structure, it is
necessary to calculate the corresponding matrix element of the quark
current between diquark states. Such calculation leads to the
emergence of the form factor $F(r)$ entering the vertex of the
diquark-gluon interaction \cite{hbar}. This form factor is expressed
through the overlap integral of the diquark wave functions.
Using the numerical diquark wave functions we find that  $F(r)$ can be
approximated  with a high accuracy by the expression \cite{hbar} 
\begin{equation}
  \label{eq:fr}
  F(r)=1-e^{-\xi r -\zeta r^2}.
\end{equation}
The values of the parameters $\xi$ and $\zeta$ for the  ground states
of the  scalar $[q,q']$ and axial vector $\{q,q'\}$
light diquarks are given in Table~\ref{tab:fcc}. 

\begin{table}
\caption{\label{tab:fcc}Parameters  $\xi$ and $\zeta$ for ground state
  light diquarks.}
\begin{ruledtabular}
\begin{tabular}{cccc}
Quark &Diquark& $\xi$  & $\zeta$  \\
content& type&(GeV)&(GeV$^2$)\\
\hline
$[u,d]$&S & 1.09 & 0.185  \\
$\{u,d\}$&A &1.185 & 0.365  \\
$[u,s]$& S & 1.23 & 0.225 \\
$\{u,s\}$& A & 1.15 & 0.325\\
$\{s,s\}$ & A& 1.13 & 0.280
\end{tabular}
\end{ruledtabular}
\end{table}

At the second step, we calculate the masses of heavy baryons as the bound
states of a heavy quark and light diquark.
For  the potential of the heavy-quark--light-diquark
interaction (\ref{dpot}) we use the expansion in $p/m_Q$. Since the
light diquark is not  heavy enough for the applicability of a $p/m_d$
expansion, it should be treated fully relativistically. 
To simplify the potential we follow the same procedure,
which was used for light quarks in a diquark,  and replace
the diquark energies $E_d(p)=\sqrt{{\bf p}^2+M_d^2}\to
E_d=(M^2-m_Q^2+M_d^2)/(2M)$ in  
Eqs.~(\ref{dpot}), (\ref{eq:jc}). This substitution makes the Fourier
transform of the potential (\ref{dpot}) local.  
At leading order in $p/m_Q$ the resulting
potential can be presented in the form:\\
for the scalar  diquark 
\begin{equation}
\label{v0s}
 V^{(0)}(r)= \hat V_{\rm Coul}(r)
+V_{\rm conf}(r),
\end{equation}
and for the axial vector  diquark
\begin{eqnarray}
\label{v0a}
 V^{(0)}(r)&= &\hat V_{\rm Coul}(r)
+V_{\rm conf}(r) +\frac{1}{M_d(E_d+M_d)}\frac1{r}\Biggl[\frac{M_d}{E_d}
\hat  V'_{\rm Coul}(r)\cr 
&&-V'_{\rm conf}(r)+\mu_d\frac{E_d+M_d}{2M_d}V'^V_{\rm conf}(r)
\Biggr]{\bf L}{\bf S}_d,
\end{eqnarray}
$$\hat V_{\rm Coul}(r)=-\frac43\alpha_s\frac{ F(r)}{r}, \qquad 
V_{\rm   conf}(r)=V^S_{\rm   conf}(r)+V^V_{\rm   conf}(r)=Ar+B,$$
 where $\hat V_{\rm Coul}(r)$ is the smeared Coulomb
potential (which accounts for the diquark structure). 
Note that both the one-gluon exchange and confining potential contribute
to the diquark spin-orbit interaction.
In this limit
the heavy baryon levels are degenerate doublets with respect to the
heavy quark spin, since the heavy quark spin-orbit and spin-spin
interactions arise only at first order in $p/m_{Q}$. Solving
Eq.~(\ref{quas}) numerically  we  get the spin-independent part of
the baryon wave function $\Psi_B$. Then the total baryon wave function 
is a product of $\Psi_B$ and the spin-dependent part $U_B$ (for
details see Eq. (43) of Ref.~\cite{dhbd}). 

The leading order degeneracy of heavy baryon states is broken by $p/m_{Q}$
corrections. 
The ground-state quark-diquark potential (\ref{dpot}) up to the second
order of the $p/m_{Q}$ expansion is given by the
following expressions: 

(a) scalar diquark
\begin{eqnarray}
\label{svcor}\!\!\!\!\!\!\!
\delta V(r)&=&\frac{1}{E_dm_{Q}}\Bigg\{{\bf p}\left[\hat V_{\rm
Coul}(r)+V^V_{\rm conf}(r)\right]{\bf p}+\hat V'_{\rm
Coul}(r)\frac{{\bf L}^2}{2r}
-\frac{1}{4}\Delta V^V_{\rm conf}(r)\cr
&&+\frac1{r}\left(\hat V'_{\rm Coul}(r)+(1+\kappa)V'^V_{\rm
    conf}(r)\right)
{\bf L}{\bf S}_Q\Bigg\}\cr
&&+\frac1{m_Q^2}\Biggl\{\frac18\Delta\left(\hat V_{\rm
Coul}(r)+V^S_{\rm conf}(r)-[1-2(1+\kappa)]V^V_{\rm conf}(r)
\right)
-\frac12{\bf p}V^S_{\rm conf}(r){\bf p}\cr
&&+\frac1{2r}\left(\hat V'_{\rm Coul}(r)-V'_{\rm conf}(r)+2(1+\kappa)
V'^V_{\rm conf}(r)\right){\bf L}{\bf S}_Q\Biggr\},
\end{eqnarray}

(b) axial vector diquark
\begin{eqnarray}
\label{avcor}\!\!\!\!\!\!\!
\delta V(r)&=&\frac{1}{E_dm_{Q}}\Bigg\{{\bf p}\left[\hat V_{\rm
Coul}(r)+V^V_{\rm conf}(r)\right]{\bf p}+\hat V'_{\rm
Coul}(r)\frac{{\bf L}^2}{2r}
-\frac{1}{4}\Delta V^V_{\rm conf}(r)\cr
&&+\frac1{r}\left(\hat V'_{\rm Coul}(r)+\frac{\mu_d}2V'^V_{\rm
    conf}(r)\right)
{\bf L}{\bf S}_d
+\frac1{r}\left(\hat V'_{\rm Coul}(r)+(1+\kappa)V'^V_{\rm
    conf}(r)\right)
{\bf L}{\bf S}_Q\cr
& & 
+\frac13\Biggl(\frac1{r}\hat V'_{\rm Coul}(r)-\hat V''_{\rm
    Coul}(r)+\frac{\mu_d}2 (1+\kappa)\left[\frac1{r}V'^V_{\rm
      conf}(r)- V''^V_{\rm conf}(r)\right]\Biggr)\cr
&&\times\left[-{\bf S}_d{\bf
    S}_Q +\frac3{r^2}({\bf S}_d{\bf r})({\bf S}_Q{\bf r})\right]
+\frac23\left[\Delta \hat V_{\rm Coul}(r)+
\frac{\mu_d}2(1+\kappa)\Delta V^V_{\rm 
conf}(r)\right]{\bf S}_{d}{\bf S}_Q\Biggr\}\cr
&&+\frac1{m_Q^2}\Biggl\{\frac18\Delta\left(\hat V_{\rm
Coul}(r)+V^S_{\rm conf}(r)-[1-2(1+\kappa)]V^V_{\rm conf}(r)
\right)
-\frac12{\bf p}V^S_{\rm conf}(r){\bf p}\cr
&&+\frac1{2r}\left(\hat V'_{\rm Coul}(r)-V'_{\rm conf}(r)+2(1+\kappa)
V'^V_{\rm conf}(r)\right){\bf L}{\bf S}_Q\Biggr\},
\end{eqnarray}
where  ${\bf L}$ is the orbital momentum, ${\bf S}_{d}$ and ${\bf
  S}_{Q}$ are the light diquark and heavy 
quark spins, respectively.
It is necessary to note that the confining vector interaction gives a
contribution to the spin-dependent part at first order of the heavy
quark expansion which is proportional to
$(1+\kappa)$ or $\mu_d$. Thus it vanishes for the chosen values
of $\kappa=-1$ and $\mu_d=0$, while the confining vector contribution to 
the spin-independent part is nonzero at this order. The first nonvanishing
contribution of the confining interaction to the heavy quark
spin-orbit part arises only at second order of the heavy quark expansion. 

Now we can calculate the mass spectra of heavy 
baryons with the account of all  corrections  of
order $p^2/m_Q^2$. For this purpose we consider 
Eq.~(\ref{quas}) with  the quasipotential
which is the sum of the leading order potentials $V^{(0)}(r)$
(\ref{v0s}) or (\ref{v0a}) and
the corrections $\delta V(r)$ (\ref{svcor}), (\ref{avcor}),
respectively. We average the resulting equation over the wave
functions of Eq.~(\ref{quas}) calculated with the leading order
potential $V^{(0)}(r)$.   
In this way we obtain the mass equation 
\begin{equation}
\label{mform}
\frac{b^2(M)}{2\mu_R}=\frac{\langle{\bf
p}^2\rangle}{2\mu_{R}}+\langle V^{(0)}(r)\rangle+\langle\delta V(r)  \rangle.
\end{equation}

It is important to note that the presence of the spin-orbit
interaction ${\bf L}{\bf S}_Q$ and of the tensor interaction
in the quark-diquark potential 
(\ref{svcor})--(\ref{avcor}) results in a mixing of states which
have the same total angular momentum $J$ and parity $P$ but different light
diquark total angular momentum  (${\bf L}+{\bf S}_d$). Such mixing is
considered along the same lines 
as in our previous calculations of the mass spectra of doubly heavy
baryons \cite{efgm}.

The calculated values of the ground state and excited baryon masses are given in
Tables~\ref{tab:lq}-\ref{tab:om} in comparison with available
experimental data \cite{pdg,cdfLambda,cdfSigma,babar2940,belle2940,belle,babar}.
In the first two 
columns we put the baryon quantum numbers and the 
state of the heavy-quark--light-diquark bound system (in usual
notations $nL$), while in the rest
columns our predictions for the masses and experimental data are
shown.

\begin{table}
\caption{\label{tab:lq} Masses of the $\Lambda_Q$ ($Q=c,b$) heavy baryons (in MeV).}
\begin{ruledtabular}
\begin{tabular}{ccccccc}
& & \multicolumn{2}{c}{\hspace{-1cm}\underline{\hspace{1.5cm}$Q=c$\hspace{1.5cm}}}\hspace{-0.8cm}& \multicolumn{3}{c}{\hspace{-1cm}\underline{\hspace{2.8cm}$Q=b$\hspace{2.8cm}}}\hspace{-1cm}\\
$I(J^P)$& $Qd$ state & $M$ & $M^{\rm exp}$ \cite{pdg}&  $M$ & $M^{\rm
  exp}$  \cite{pdg}& $M^{\rm exp}$ \cite{cdfLambda}\\
\hline
$0(\frac12^+)$& $1S$ & 2297 & 2286.46(14) & 5622 & 5624(9)&5619.7(2.4)\\
$0(\frac12^-)$& $1P$ & 2598 & 2595.4(6) & 5930 & \\
$0(\frac32^-)$& $1P$ & 2628 & 2628.1(6) & 5947 & \\
$0(\frac12^+)$& $2S$ & 2772 & 2766.6(2.4)?& 6086 & \\
$0(\frac32^+)$& $1D$ & 2874 &           & 6189 & \\
$0(\frac52^+)$& $1D$ & 2883 & 2882.5(2.2)?& 6197 & \\
$0(\frac12^-)$& $2P$ & 3017 &           & 6328 & \\
$0(\frac32^-)$& $2P$ & 3034 &           & 6337 & \\
$0(\frac52^-)$& $1F$ & 3061 &           & 6401 & \\
$0(\frac72^-)$& $1F$ & 3057 &           & 6405 & \\
$0(\frac12^+)$& $3S$ & 3150 &           & 6465 & \\
$0(\frac32^+)$& $2D$ & 3262 &           & 6540 & \\
$0(\frac52^+)$& $2D$ & 3268 &           & 6548& \\
\end{tabular}
\end{ruledtabular}
\end{table}

\begin{table}
\caption{\label{tab:sq} Masses of the $\Sigma_Q$ ($Q=c,b$) heavy baryons (in MeV).}
\begin{ruledtabular}
\begin{tabular}{ccccccccc}
& & \multicolumn{4}{c}{\hspace{-0.8cm}\underline{\hspace{3.4cm}$Q=c$\hspace{3.4cm}}}\hspace{-0.8cm}& \multicolumn{3}{c}{\hspace{-1cm}\underline{\hspace{1.8cm}$Q=b$\hspace{1.8cm}}}\hspace{-1cm}\\
$I(J^P)$& $Qd$ state & $M$ & $M^{\rm exp}$  \cite{pdg} & $M^{\rm exp}$
\cite{babar2940}& $M^{\rm exp}$  \cite{belle2940} &  $M$ &
$M^{\rm exp}$ \cite{cdfSigma}&
$M^{\rm exp}$ \cite{cdfSigma}\\
\hline
$1(\frac12^+)$& $1S$ & 2439 & 2453.76(18)&      &   & 5805 & 5807.5$^*$& 5815.2$^\dag$\\
$1(\frac32^+)$& $1S$ & 2518 & 2518.0(5)  &      &   & 5834 & 5829.0$^*$& 5836.7$^\dag$\\
$1(\frac12^-)$& $1P$ & 2805 &            &      &   & 6122 &     \\
$1(\frac12^-)$& $1P$ & 2795 &            &      &   & 6108 &    \\
$1(\frac32^-)$& $1P$ & 2799 & 2802($^4_7$)&     &   & 6106 & \\
$1(\frac32^-)$& $1P$ & 2761 & 2766.6(2.4)?&     &   & 6076 & \\
$1(\frac52^-)$& $1P$ & 2790 &            &      &   & 6083 & \\
$1(\frac12^+)$& $2S$ & 2864 &            &      &   & 6202 & \\
$1(\frac32^+)$& $2S$ & 2912 &            &2939.8(2.3)?&2938($^3_5$)?& 6222 &\\
$1(\frac12^+)$& $1D$ & 3014 &            &      &    & 6300 & \\
$1(\frac32^+)$& $1D$ & 3005 &            &      &    & 6287 & \\
$1(\frac32^+)$& $1D$ & 3010 &            &      &    & 6291 & \\
$1(\frac52^+)$& $1D$ & 3001 &            &      &    & 6279 & \\
$1(\frac52^+)$& $1D$ & 2960 &            &      &    & 6248 & \\
$1(\frac72^+)$& $1D$ & 3015 &            &      &    & 6262 & \\
$1(\frac12^-)$& $2P$ & 3186 &            &      &    & 6411 & \\
$1(\frac12^-)$& $2P$ & 3176 &            &      &    & 6401 & \\
$1(\frac32^-)$& $2P$ & 3180 &            &      &    & 6400 & \\
$1(\frac32^-)$& $2P$ & 3147 &            &      &    & 6379 & \\
$1(\frac52^-)$& $2P$ & 3167 &            &      &    & 6383 & \\
\end{tabular}
\end{ruledtabular}
$^*$ data for $\Sigma_b^{(*)+}$, experimental errors are given in the text\\
$^\dag$ data for $\Sigma_b^{(*)-}$, experimental errors are given in the text
\end{table}

\begin{table}
\caption{\label{tab:xs} Masses of the $\Xi_Q$ ($Q=c,b$) heavy baryons
  with scalar diquark (in MeV).}
\begin{ruledtabular}
\begin{tabular}{ccccc}
& & \multicolumn{2}{c}{\hspace{-1cm}\underline{\hspace{2.2cm}$Q=c$\hspace{2.2cm}}}\hspace{-0.8cm}& \multicolumn{1}{c}{\underline{\hspace{0.2cm}$Q=b$\hspace{0.2cm}}}\\
$I(J^P)$& $Qd$ state & $M$ & $M^{\rm exp}$ \cite{pdg} & $M$ \\
\hline
$\frac12(\frac12^+)$& $1S$ & 2481 & 2471.0(4)   & 5812 \\
$\frac12(\frac12^-)$& $1P$ & 2801 & 2791.9(3.3) & 6119  \\
$\frac12(\frac32^-)$& $1P$ & 2820 & 2818.2(2.1) & 6130  \\
$\frac12(\frac12^+)$& $2S$ & 2923 &             & 6264  \\
$\frac12(\frac32^+)$& $1D$ & 3030 &             & 6359  \\
$\frac12(\frac52^+)$& $1D$ & 3042 &             & 6365  \\
$\frac12(\frac12^-)$& $2P$ & 3186 &             & 6492  \\
$\frac12(\frac32^-)$& $2P$ & 3199 &             & 6494  \\
$\frac12(\frac52^-)$& $1F$ & 3219 &             & 6555  \\
$\frac12(\frac72^-)$& $1F$ & 3208 &             & 6558  \\
$\frac12(\frac12^+)$& $3S$ & 3313 &             & 6618  \\
$\frac12(\frac32^+)$& $2D$ & 3411 &             & 6688  \\
$\frac12(\frac52^+)$& $2D$ & 3413 &             & 6692 \\
\end{tabular}
\end{ruledtabular}
\end{table}

\begin{table}
\caption{\label{tab:xv} Masses of the $\Xi_Q$ ($Q=c,b$) heavy baryons
  with axial vector diquark (in MeV).}
\begin{ruledtabular}
\begin{tabular}{ccccccc}
& & \multicolumn{4}{c}{\hspace{-1cm}\underline{\hspace{4.4cm}$Q=c$\hspace{4.4cm}}}\hspace{-0.8cm}& {\underline{\hspace{0.2cm}$Q=b$\hspace{0.2cm}}}\\
$I(J^P)$& $Qd$ state & $M$ & $M^{\rm exp}$  \cite{pdg} & $M^{\rm exp}$  \cite{belle}& $M^{\rm exp}$  \cite{babar}&  $M$ \\
\hline
$\frac12(\frac12^+)$& $1S$ & 2578 & 2578.0(2.9) & & & 5937 \\
$\frac12(\frac32^+)$& $1S$ & 2654 & 2646.1(1.2) & & & 5963 \\
$\frac12(\frac12^-)$& $1P$ & 2934 &           & &   & 6249    \\
$\frac12(\frac12^-)$& $1P$ & 2928 &           & &   & 6238     \\
$\frac12(\frac32^-)$& $1P$ & 2931 &           & &   & 6237  \\
$\frac12(\frac32^-)$& $1P$ & 2900 &           & &   & 6212  \\
$\frac12(\frac52^-)$& $1P$ & 2921 &           & &   & 6218  \\
$\frac12(\frac12^+)$& $2S$ & 2984 & &2978.5(4.1)&2967.1(2.9) & 6327 \\
$\frac12(\frac32^+)$& $2S$ & 3035 &           & &   & 6341 \\
$\frac12(\frac12^+)$& $1D$ & 3132 &           & &   & 6420  \\
$\frac12(\frac32^+)$& $1D$ & 3127 &           & &   & 6410 \\
$\frac12(\frac32^+)$& $1D$ & 3131 &           & &   & 6412 \\
$\frac12(\frac52^+)$& $1D$ & 3123 &           & &   & 6403 \\
$\frac12(\frac52^+)$& $1D$ & 3087 & &3082.8(3.3)&3076.4(1.0) & 6377 \\
$\frac12(\frac72^+)$& $1D$ & 3136 &           & &   & 6390 \\
$\frac12(\frac12^-)$& $2P$ & 3300 &           & &   & 6527 \\
$\frac12(\frac12^-)$& $2P$ & 3294 &           & &   & 6519 \\
$\frac12(\frac32^-)$& $2P$ & 3296 &           & &   & 6518 \\
$\frac12(\frac32^-)$& $2P$ & 3269 &           & &   & 6500  \\
$\frac12(\frac52^-)$& $2P$ & 3282 &           & &   & 6504  \\
\end{tabular}
\end{ruledtabular}
\end{table}

\begin{table}
\caption{\label{tab:om} Masses of the $\Omega_Q$ ($Q=c,b$) heavy baryons (in MeV).}
\begin{ruledtabular}
\begin{tabular}{cccccc}
& & \multicolumn{3}{c}{\hspace{-1.2cm}\underline{\hspace{3.4cm}$Q=c$\hspace{3.4cm}}}\hspace{-0.8cm}& \multicolumn{1}{c}{\underline{\hspace{0.2cm}$Q=b$\hspace{0.2cm}}}\\
$I(J^P)$& $Qd$ state & $M$ & $M^{\rm exp}$  \cite{pdg}&$M^{\rm exp}$  \cite{babaromega} &  $M$ \\
\hline
$0(\frac12^+)$& $1S$ & 2698 & 2697.5(2.6)& & 6065\\
$0(\frac32^+)$& $1S$ & 2768 & &2768.3(3.0) & 6088\\
$0(\frac12^-)$& $1P$ & 3025 &           &  & 6361     \\
$0(\frac12^-)$& $1P$ & 3020 &           &  & 6352   \\
$0(\frac32^-)$& $1P$ & 3026 &           &  & 6351 \\
$0(\frac32^-)$& $1P$ & 2998 &           &  & 6330\\
$0(\frac52^-)$& $1P$ & 3022 &           &  & 6336  \\
$0(\frac12^+)$& $2S$ & 3065 &           &  & 6440 \\
$0(\frac32^+)$& $2S$ & 3119 &           &  & 6454 \\
$0(\frac12^+)$& $1D$ & 3222 &           &  & 6526 \\
$0(\frac32^+)$& $1D$ & 3215 &           &  & 6518  \\
$0(\frac32^+)$& $1D$ & 3217 &           &  & 6520 \\
$0(\frac52^+)$& $1D$ & 3218 &           &  & 6512 \\
$0(\frac52^+)$& $1D$ & 3187 &           &  & 6490  \\
$0(\frac72^+)$& $1D$ & 3237 &           &  & 6502  \\
$0(\frac12^-)$& $2P$ & 3376 &           &  & 6630 \\
$0(\frac12^-)$& $2P$ & 3371 &           &  & 6624 \\
$0(\frac32^-)$& $2P$ & 3374 &           &  & 6623 \\
$0(\frac32^-)$& $2P$ & 3350 &           &  & 6608 \\
$0(\frac52^-)$& $2P$ & 3365 &           &  & 6611 \\
\end{tabular}
\end{ruledtabular}
\end{table}

At present the best experimentally studied quantities are the mass
spectra of the
$\Lambda_Q$ and $\Sigma_Q$ baryons, which contain the light scalar or
axial vector diquarks, respectively. They are presented in
Tables~\ref{tab:lq}, \ref{tab:sq}. Masses of the ground
states are measured both for charmed and bottom
$\Lambda_Q$ and $\Sigma_Q$ baryons. Note that the masses of the
ground state $\Sigma_b$ and $\Sigma^*_b$ baryons were first reported very recently by
CDF \cite{cdfSigma}: 
$M_{\Sigma_b^+}=5807.5^{+1.9}_{-2.2}\pm1.7$~MeV,
$M_{\Sigma_b^-}=5815.2^{+1.0}_{-0.9}\pm1.7$~MeV,
$M_{\Sigma_b^{*+}}=5829.0^{+1.6}_{-1.7}\pm1.7$~MeV,
$M_{\Sigma_b^{*-}}=5836.7^{+2.0}_{-1.8}\pm1.7$~MeV. 
CDF also significantly improved the precision of
the $\Lambda_b$ mass \cite{cdfLambda}. For charmed baryons the masses
of several excited states are also known. It is important to emphasize that
the $J^P$ quantum numbers for most excited heavy baryons have not
been determined experimentally, but are assigned by PDG on the basis of
quark model predictions. For some excited charm
baryons such as the $\Lambda_c(2765)$, $\Lambda_c(2880)$ and
$\Lambda_c(2940)$  it is even not known if they are excitations of the
$\Lambda_c$ or $\Sigma_c$.~\footnote{In Tables~\ref{tab:lq},
  \ref{tab:sq}, \ref{tab:comp} we mark with ? the states which interpretation is
  ambiguous.} Our calculations show that the
$\Lambda_c(2765)$ can be either the first radial (2$S$) excitation of
the $\Lambda_c$ with  $J^P=\frac12^+$ containing the light
scalar diquark or the first orbital excitation (1$P$) of the 
$\Sigma_c$  with $J^P=\frac32^-$  containing the light axial
vector diquark. The $\Lambda_c(2880)$ baryon in our model is well
described by the second orbital (1$D$) excitation of the $\Lambda_c$
with  $J^P=\frac52^+$ in 
agreement with the recent spin assignment  \cite{belle2940}
based on the analysis of angular distributions in the  decays
$\Lambda_c(2880)^+\to\Sigma_c(2455)^{0,++}\pi^{+,-}$. Our model
suggests that the   charmed baryon $\Lambda_c(2940)$, recently
discovered  by BaBar\cite{babar2940}  and then also 
confirmed by Belle \cite{belle2940}, could be the first radial (2$S$) excitation of
the $\Sigma_c$ with $J^P=\frac32^+$ which mass is predicted slightly
below the experimental value. If this state  proves to be an excited
$\Lambda_c$, for which we have no candidates around 2940 MeV, then it
will indicate that excitations inside the diquark should be also
considered.~\footnote{The $\Lambda_c$ baryon with the first
orbital excitation of the diquark is expected to have a mass in this
region.}   The $\Sigma_c(2800)$ baryon can be identified in our model
with one of the orbital (1$P$) excitations of the $\Sigma_c$
with $J^P=\frac12^-, \frac32^-$ or $\frac52^-$ which predicted mass
differences are  less than 15 MeV. Thus masses of all these states
are  compatible with the experimental value within errors.       
  
Mass spectra of the $\Xi_Q$ baryons with the scalar and axial vector light
($qs$) diquarks are given in Tables~\ref{tab:xs}, \ref{tab:xv}. Experimental
data here are available only for charm-strange baryons. We can identify the
$\Xi_c(2790)$ and $\Xi_c(2815)$ with the first orbital (1$P$)
excitations of the $\Xi_c$  with $J^P=\frac12^-$ and $J^P=\frac32^-$,
respectively, containing the light scalar diquark, which is
in agreement with the PDG \cite{pdg} assignment. Recently Belle
\cite{belle} reported the first observation of two baryons
$\Xi_{cx}(2980)$ and $\Xi_{cx}(3077)$, which existence was also 
confirmed by BaBar \cite{babar}. The $\Xi_{cx}(2980)$ can be interpreted
in our model as the first radial (2$S$) excitation of the $\Xi_c$  with
$J^P=\frac12^+$ containing the light axial vector diquark. On the other hand the
$\Xi_{cx}(3077)$  corresponds to the second orbital (1$D$) excitation
in this system with $J^P=\frac52^+$.   

For the $\Omega_Q$ baryons only masses of the ground-state charmed baryons
are known. The $\Omega^*_c$ baryon was very recently discovered by
BaBar \cite{babaromega}. The measured mass difference of the
$\Omega^*_c$ and  $\Omega_c$ baryons of ($70.8\pm1.0\pm1.1$)~MeV is
in very good agreement with the prediction of our model 70~MeV \cite{hbar}.

Our predictions for the heavy baryon mass spectra can be also compared 
with results of other calculations, e.g. \cite{ci,mmmp,gvv}. In
Ref.~\cite{ci} the variational approach is used to solve the three-body
problem in the relativized quark model with the QCD motivated quark
potential. Authors of Ref.~\cite{mmmp} calculate the mass spectra of
charmed baryons within a relativistic quark model based on the
Salpeter equation with a potential containing both the confining
potential and instanton induced interactions. In Ref.~\cite{gvv}
the three-quark problem is solved by means of the Faddeev method in
momentum space  with the quark-quark interaction consisting of
the one-gluon exchange, confinement and boson exchange
potentials. All these approaches are three-body ones and thus they predict
the mass spectra of excited heavy baryons with significantly more
levels than we get in our model, since we use the quark-diquark
approximation. The comparison given in Table~\ref{tab:comp} shows that
our predictions agree with 
experiment in most cases better than the results of the above
mentioned approaches.  The most clear example is our prediction
\cite{hbar} for the masses of the $\Omega_c^*$ and $\Sigma_b$,
$\Sigma_b^*$, which agree with experiment with high accuracy.
The accurate predictions for the $\Sigma_b$ and $\Sigma_b^*$ masses
are also given in Ref.~\cite{lipkin}.  

\begin{table}
\caption{\label{tab:comp} Comparison of theoretical predictions for
  masses (in MeV) of heavy baryons (for $J=\frac12,\frac32$) with
  experimental data.\protect\footnote{
  Only central values of measured masses are given. Experimental errors can be found in Tables~\ref{tab:lq}-\ref{tab:om}.}}
\begin{ruledtabular}
\begin{tabular}{ccccccccccc}
$J^P$&exp.&our & \cite{ci} &\cite{mmmp} &\cite{gvv}&exp.&our &
\cite{ci} &\cite{mmmp} &\cite{gvv}\\
\hline
&
\multicolumn{5}{c}{\hspace{-1.2cm}\underline{\hspace{3.4cm}$\Lambda_c$\hspace{3.4cm}}}\hspace{-0.8cm}&\multicolumn{5}{c}{\hspace{-1.2cm}\underline{\hspace{3.4cm}$\Sigma_c$\hspace{3.4cm}}}\hspace{-0.8cm}\\
$\frac12^+$&2286 &2297&2265&2272&2292 &2454&2439&2440&2459&2448\\
$\frac12^+$&2766?&2772&2775&2769&2669 &    &2864&2890&2947&2793\\
$\frac32^+$&     &2874&2910&2848&2906 &2518&2518&2495&2539&2505\\
$\frac32^+$&     &3262&3035&3100&3061 &    &2912&2985&3010&2825\\
$\frac12^-$&2595 &2598&2630&2594&2559&2802?&2795&2765&2769&2706\\
$\frac12^-$&     &3017&2780&2853&2779&2802?&2805&2770&2817&2791\\
$\frac32^-$&2628 &2628&2640&2586&2559&2766?&2761&2770&2799&2706\\
$\frac32^-$&     &3034&2840&2874&2779&2802?&2799&2805&2815&2791\\
&\multicolumn{5}{c}{\hspace{-1.2cm}\underline{\hspace{3.4cm}$\Xi_c$\hspace{3.4cm}}}\hspace{-0.8cm}&\multicolumn{5}{c}{\hspace{-1.2cm}\underline{\hspace{3.4cm}$\Omega_c$\hspace{3.4cm}}}\hspace{-0.8cm}\\
$\frac12^+$&2471&2481& &2469&2496& 2698&2698& &2688&2701\\
$\frac12^+$&2578&2578& &2595&2574&     &3065& &3169&3044\\
$\frac32^+$&2646&2654& &2651&2633& 2768&2768& &2721&2759\\
$\frac32^+$&    &3030& &    &2951&     &3119& &    &3080\\
$\frac12^-$&2792&2801& &2769&2749&     &3020& &    &2959\\
$\frac12^-$&    &2928& &    &2829&     &3025& &    &3029\\
$\frac32^-$&2818&2820& &2771&2749&     &2998& &    &2959\\
$\frac32^-$&    &2900& &    &2829&     &3026& &    &3029\\
&
\multicolumn{5}{c}{\hspace{-1.2cm}\underline{\hspace{3.4cm}$\Lambda_b$\hspace{3.4cm}}}\hspace{-0.8cm}&\multicolumn{5}{c}{\hspace{-1.2cm}\underline{\hspace{3.4cm}$\Sigma_b$\hspace{3.4cm}}}\hspace{-0.8cm}\\
$\frac12^+$&5620&5622&5585& &5624& 5808&5805&5795& &5789\\
$\frac32^+$&    &6189&6145& &6246& 5829&5834&5805& &5844\\
$\frac12^-$&    &5930&5912& &5890&     &6108&6070& &6039\\
$\frac32^-$&    &5947&5920& &5890&     &6076&6070& &6039\\
&\multicolumn{5}{c}{\hspace{-1.2cm}\underline{\hspace{3.4cm}$\Xi_b$\hspace{3.4cm}}}\hspace{-0.8cm}&\multicolumn{5}{c}{\hspace{-1.2cm}\underline{\hspace{3.4cm}$\Omega_b$\hspace{3.4cm}}}\hspace{-0.8cm}\\
$\frac12^+$&    &5812& & &5825&   &6065& & &6037\\
$\frac32^+$&    &5963& & &5967&   &6088& & &6090\\
$\frac12^-$&    &6119& & &6076&   &6352& & &6278\\
$\frac32^-$&    &6130& & &6076&   &6330& & &6278
\end{tabular}
\end{ruledtabular}
\end{table}

In conclusion we emphasize that, in calculating the heavy baryon
masses, we do not use any free adjustable parameters, thus all obtained
results are pure predictions. Indeed, the values of all
parameters of the model (including quark masses and parameters of
the quark potential) were fixed in our previous considerations of
meson properties. Note 
that the light diquark in our approach is not considered as a
point-like object. Instead we use its wave functions to calculate  
diquark-gluon interaction form factors and thus take into account
the finite (and relatively large) 
size of the light diquark. The other important advantage of our
model is the completely relativistic treatment of the light quarks
in the diquark and of the light diquark in the heavy baryon. We use the $v/c$
expansion only for heavy ($b$ and $c$) quarks.     

We find that all presently available experimental data for the ground and excited
states of heavy baryons can be accommodated in the picture
treating a heavy baryon as the bound system of the light diquark
and heavy quark, experiencing orbital and radial excitations only
between these constituents.   

The obtained wave functions of the ground-state and excited heavy
baryons can be used for calculations of the semileptonic and nonleptonic
weak decays and of the one-pion transitions between excited and
ground states. The heavy-to-heavy semileptonic decays of bottom
baryons to charmed baryons were already studied by us 
in Ref.~\cite{seml}. For the calculation of the heavy-to-light 
semileptonic decays the light baryon wave functions are
necessary. The application of a simple quark-diaquark approximation for
light baryons is controversial and thus more sophisticated methods
should be used.  

Note added: After this letter was submitted for publication the D0
Collaboration \cite{d0xib} reported the discovery of the $\Xi_b^-$
baryon with the 
mass $M_{\Xi_b}=5774\pm11\pm15$~MeV. The CDF Collaboration
\cite{cdfxib} confirmed this observation and gave the more
precise value $M_{\Xi_b}=5792.9\pm2.5\pm1.7$~MeV. Our model prediction
$M_{\Xi_b}=5812$~MeV is in a reasonable agreement with these new
data.  The BaBar Collaboration \cite{babarxic} announced
observation of two new charmed baryons $\Xi_c(3055)$ with the mass
$M=3054.2\pm1.2\pm0.5$~MeV and $\Xi_c(3123)$ with the mass
$M=3122.9\pm1.3\pm0.3$~MeV. These states can be interpreted in our
model as the second orbital ($1D$) excitations of the $\Xi_c$  with
$J^P=\frac52^+$ containing scalar and axial vector diquarks,
respectively. Their predicted masses are 3042~MeV and 3123~MeV.         

The authors are grateful to M. Ivanov, M. M\"uller-Preussker, J. Rosner
and V. Savrin  for support and discussions.  Two of us
(R.N.F. and V.O.G.)  were supported in part by the {\it Deutsche
Forschungsgemeinschaft} under contract Eb 139/2-4 and by the {\it Russian
Foundation for Basic Research} under Grant No.05-02-16243.

\end{document}